\documentclass[12pt]{iopart}
\newcommand{\eqref}[1]{\mbox{(\ref{#1})}}
\usepackage[utf8]{inputenc}
\usepackage{iopams}
\begin {document}

\title[Averaging in cosmology based on Cartan scalars]{Averaging in cosmology based on Cartan scalars}

\author{P Ka\v spar, O Sv\'itek}
\address{Institute of Theoretical Physics, Faculty of Mathematics and Physics, Charles University in Prague, V Holesovickach 2, 180 00 Prague, Czech Republic}
\ead{petrkaspar@atlas.cz, ota@matfyz.cz}

\begin{abstract}
We present a new approach for averaging in general relativity and cosmology. After a short review of the theory originally taken from the equivalence problem, we consider two ways how to deal with averaging based on Cartan scalars. We apply the theory for two different LTB models. In the first one, correlation term behaves as a positive cosmological constant, in the second example leading correlation term behaves like spatial curvature. We also show nontriviality of averaging for linearized monochromatic gravitational wave.
\end{abstract}

\section {Introduction}
\indent \indent In general relativity and cosmology, we are often dealing with spacetimes that have many symmetries. We can justify this step by choosing some particular length scale and claim, that our simple spacetime is the average of some more realistic model. The main motivation for the averaging comes from cosmology. Gravity is well tested within our solar system. On cosmological scales we do not need to know the details about fluctuating gravitational field. In order to obtain a "macroscopic" theory of gravity we should perform averaging of Einstein equations. These equations are strongly nonlinear, so if we want to use averaged metric, we have to add a correlation term which does not need to satisfy usual energy conditions and can act as a dark energy. The problem is that averaging involves integration of the tensor field on the curved manifold and this operation is not well defined. 
\newline \indent The most popular approach to averaging is scalar averaging and investigation of so called Buchert equations~\cite{Buchert1},~\cite{Buchert2}, where only scalar part of the Einstein equations is averaged (see \cite{Buchert3} for a recent review). All Einstein equations are averaged in the context of Macroscopic Gravity~\cite{Zal1},~\cite{Zal2}, at the same time the Cartan structure equations which describe geometry of spacetime are averaged. Theorem about isometric embedding of a 2-sphere into Euclidian space is applied for averaging by Korzy\~{n}ski~\cite{Korz}. In~\cite{Brannlund} the Weitzenb{\"o}ck connection for parallel transport is used for the definition of average value of tensor field. 
\newline \indent The theory of Cartan scalars was developed in order to decide if two spacetimes are locally equivalent~\cite{Cartan},~\cite{Karl1}. We can use this theory for local characterization of given spacetime. Then, inspired by method given by Coley~\cite{scalars}, who investigated averaged scalar invariants constructed from the Riemann tensor and finite number of its covariant derivatives, we average the left hand side of Einstein equations (which contain finite number of the Cartan scalars if rewritten in tetrad form) and we give the prescription for the computation of the correlation term.     
\newline \indent In the first section, we review the theory of Cartan scalars, then after a short introduction of LTB spacetime, we give two different examples of averaging by Cartan scalars.  The first one utilizes approximation for the areal function $R(t,r)$. In the second example we investigate backreaction for the LTB metric given by Biswas, Mansouri and Notari~\cite{Biswas}. Then we consider averaged linearized monochromatic gravitational wave and we will end with conclusion.

\section {Cartan scalars}
\indent \indent If we want to specify the geometry of spacetime, we are allowed to choose the $\frac{n(n+1)}{2}$ components of the metric tensor. There also exists another possibility. It can be shown that the tetrad projection of Riemann tensor and the finite number of its covariant derivatives (called Cartan scalars) completely (locally) specify the geometry of Riemannian manifold~\cite{Cartan}. Cartan scalars are true scalars on the bundle of frames $F({\cal M})$ but if we fix the tetrad, they behave as scalars on the manifold as well. Because it is still not clear how to unambiguously average a metric tensor, there exists a possibility to describe the geometry with Cartan scalars and average them (which is straightforward in the case of scalars).
\newline \indent There exists another advantage within this formalism. The left hand side of the Einstein equations can be rewritten in the tetrad form, so it consists of the finite sum of Cartan scalars. Using Cartan scalars we can average not only the spacetime geometry but also the left hand side of the Einstein equations. From the Cartan scalars we can easily read off a dimension of an isometry group and we can obtain an algebra of the Killing vectors~\cite{Karl3}. 

\indent We will review the construction of the Cartan scalars~\cite{Karl1},~\cite{curv},~\cite{find}. Let ${\cal M}$ be an n-dimensional differentiable manifold with a metric
\begin{equation}
\mathbf{g}=\eta_{ij}\bomega^i \otimes \bomega^j,
\end{equation}
$\eta_{ij}$ is a constant symmetric matrix and $\bomega^i$, i=1,2...,n form a basis of the cotangent space at the point $x^\mu$. The tetrad (frame) $\bomega^i$ is for a given $\mathbf{g}$ and $\eta_{ij}$ fixed up to the generalized rotations.
\begin{equation}
\bomega^i=\omega^i_\nu (x^\mu,\xi^\Upsilon ) \mathbf{d}x^\nu, 
\end{equation}
$\xi^\Upsilon$, $\Upsilon$=1,...,$\frac{1}{2}n(n-1)$, denotes the coordinates of an orthogonal group. For simplicity we will define all geometrical objects on the enlarged $\frac{1}{2} n(n+1)$ dimensional space - the bundle of frames $F({\cal M})$. $F({\cal M})$ is locally isomorphic to the Cartesian product of an open set on the manifold (spacetime) and the orthogonal (Lorentz) group $G$ - it means that in every point $x^\mu$ there exists a fiber with coordinates $\xi^\Upsilon$. In the following we will use an enlarged exterior derivative in the form $\mathbf{d}=\mathbf{d}_x+\mathbf{d}_{\xi}$. Cartan structure equations read
\begin{equation}
\mathbf{d}\bomega^i=\bomega^j \wedge \bomega^i_{\,\,\,j},
\label{C1}
\end{equation}
\begin{equation}
\mathbf{d}\bomega^i_{\,\,\,j}=-\bomega^i_{\,\,\,k} \wedge \bomega^k_{\,\,\,j} + \frac{1}{2} R^i_{\,\,\,jkl} \bomega^k \wedge \bomega^l.
\end{equation}
with a condition
\begin{equation}
\eta_{ik}\bomega^k_{\,\,\,j}+\eta_{jk}\bomega^k_{\,\,\,i}=0.
\end{equation}
From the first equation we can compute the connection 1-form $\bomega^i_{\,\,\,j}$, next equation serves as a definition of the curvature tensor $R^i_{\,\,\,jkl}$. To generate covariant derivatives of the Riemann tensor, we repeatedly apply an exterior derivative. 
\begin{eqnarray}
\mathbf{d}R_{ijkl} &=& R_{mjkl}\bomega^m_{\,\,\,i} + R_{imkl} \bomega^m_{\,\,\,j} + R_{ijml} \bomega^m_{\,\,\,k}  + R_{ijkm} \bomega^m_{\,\,\,l} + R_{ijkl;m}\bomega^m,\nonumber\\
\mathbf{d}R_{ijkl;n} &=& R_{mjkl;n}\bomega^m_{\,\,\,i} + R_{imkl;n}\bomega^m_{\,\,\,i} + ...+ R_{ijkl;nm}\bomega^m,\nonumber\\
&.&\label{kovderR}\\
&.&\nonumber\\
&.&\nonumber
\end{eqnarray}

\indent Let $R^p$ denotes the set $\left\{R_{ijkm}, R_{ijkm;n_1},...,R_{ijkm;n_1...n_p}\right\}$ where $p$ is such that $R^{p+1}$ contains no element that is functionally independent of the elements in $R^p$. Two functions $f$ and $g$ are functionally independent if the one forms $\mathbf{d}f$ and $\mathbf{d}g$ are linearly independent. Then the set $R^{p+1}$ characterizes the geometry completely and its elements are called Cartan scalars. There exists an algorithmic way how to compute Cartan scalars~\cite{Karl2}. It uses standard form of the Riemann tensor that can be found by the Petrov and Segre algorithm (and its generalization for tensors with more indices). However, the tetrad does not need to be fixed completely. There exist some degrees of freedom which can nontrivially transform the components of other tensors, but the Cartan scalars remain fixed. This property allows us to integrate Cartan scalars over some domain ${\cal D}\subset {\cal M}$ as we will see later. 
\newline \indent If we want to specify the geometry of spacetime, we are allowed to choose the $\frac{n(n+1)}{2}$ components of the metric tensor, which satisfy the Einstein equations. If we want to use the Cartan scalars instead, there must exist some algebraic and differential equations that they have to fulfill. In other words from a given set $R^{p+1}$ we have to find the conditions in order to be able to construct one form $\bomega^i$, which satisfies the equations~\eqref{C1} -~\eqref{kovderR}. These constraints should be respected also by the averaged Cartan scalars. 
\newline \indent To see explicitly the form of the constraints it is easier to rewrite equations~\eqref{C1} -~\eqref{kovderR} in a more compact way. The connection one form is defined on the bundle of frames $F({\cal M})$ as
\begin{equation}
\label{tij}
\bomega^i_{\,\,\,j}=\gamma^i_{\,\,\,jk}\bomega^k+\btau^i_{\,\,\,j}, 
\end{equation}
where  $\btau^i_{\,\,\,j}=\tau^i_{\,\,\,j \Upsilon} \mathbf{d}\xi^\Upsilon$ generates the orthogonal group and $\gamma^i_{\,\,\,jk}$ are the Ricci rotation coefficients. It means that $\bomega^i_{\,\,\,j}$ and $\bomega^k$ are independent objects on $F({\cal M})$ and we can denote them collectively as $\left\{ \bomega^I \right\}\equiv \left\{\bomega^i, \bomega^i_j \right\}$, $I$=1,2,...$\frac{1}{2}n(n+1)$. Cartan structure equations can be rewritten into the simple form
\begin{equation}
\mathbf{d}\bomega^I=\frac{1}{2} C^I_{\,\,JK} \bomega^J \wedge \bomega^K.
\label{CartC}
\end{equation}
$C^I_{\,\,JK}$ essentially represent the Riemann tensor on $F({\cal M})$. We will denote a maximal set of the functionally independent objects in $R^p$ as ${I^\alpha}$, $\alpha$=1,...,k $\leq$ $\frac{1}{2}$n(n+1), which can be thought of as the coordinates on the bundle of frames. It means that all objects in $R^{p+1}$ are functions of ${I^\alpha}$ only. By applying an exterior derivative we will obtain an analog of the equation~\eqref{kovderR}
\begin{eqnarray}
\mathbf{d}C^I_{\,\,JK}&=&C^I_{\,\,JK,\alpha}\mathbf{d}I^\alpha \equiv C^I_{\,\,JK,\alpha}I^\alpha_{\,\,\,|L}\bomega^L \equiv C^I_{\,\,JK|L}\bomega^L, \nonumber\\
\mathbf{d}C^I_{\,\,JK|L}&=&C^I_{\,\,JK|LM}\bomega^M,\nonumber\\
&.&\\
&.&\nonumber\\
&.&\nonumber
\end{eqnarray} 
Symbol $|$ here denotes the derivative with respect to the vector field dual to the 1-form $\bomega^L$ and similarly symbol "," represents the derivative with respect to the vector field dual to $\mathbf{d}I^\alpha$. We can see from the above equations that $R^{p+1}$ can be constructed from the set $\left\{C^I_{JK},I^\alpha_{\,\,\,|L}\right\}$. The constraints that have to be satisfied then read
\begin{eqnarray}
I^\alpha_{\,\,\,|K,\beta} I^\beta_{\,\,\,|J} - I^\alpha_{\,\,\,|J,\beta}I^\beta_{\,\,\,|K} +I^\alpha_{\,\,\,|L} C^L_{\,\,JK}&=&0, \nonumber\\
C^P_{\,\,[JK|L]}+C^P_{\,\,M[K}C^M_{\,\,LJ]}&=&0\label{vazba}
\end{eqnarray}

\section {Averaging Cartan scalars}
\indent \indent Let us suppose that we have a given manifold ${\cal M}$ characterized by the set of scalar functions $R^{p+1}$ and a given domain ${\cal D}$. We would like to obtain a new manifold $\left\langle  {\cal M}\right\rangle$  - identical as a set but with a smooth metric structure, which would not recognize quickly fluctuating inhomogeneities of the gravitational field. The naive approach would consist of the integration of the scalar function $f \in R^{p+1}$ according to the rule
\begin{equation}
\left\langle f \right\rangle (x) = \frac{1}{V_{\cal D}} \int\limits_{\cal D}  {f\left( {x+x'} \right)} d^N x',
\label{stred}
\end{equation}
where $d^N x$ is an invariant metric volume element. Following this rule we would obtain a new set $\left\langle {R^{p+1}} \right\rangle$. The problem is that the elements of $\left\langle {R^{p+1}} \right\rangle$ wouldn't satisfy the constraints (which can be written as~\eqref{vazba}) because of the nonlinearity of the equations.
\newline \indent We will deal with the problem in a similar way as Coley did~\cite{scalars}. First we will restrict to the smallest possible set of independent functions $R'^{p+1} \subseteq R^{p+1}$ (with the help of the constraints it would be possible to generate the whole set $R^{p+1}$) and proceed with averaging of $R'^{p+1}$. We will obtain a new set $\left\langle {R'^{p+1}} \right\rangle$. In the next step we have to suppose, that the constraints will have the same form (they are not modified by correlation terms) and as a result we can generate the whole set $\left\langle  {R^{p+1}} \right\rangle$ from $\left\langle {R'^{p+1}} \right\rangle$. The theory then guarantees that there exists the metric tensor $\left\langle g_{\mu \nu} \right\rangle$ (or equivalently the 1-forms $\left\langle \bomega^i \right\rangle$). With a help of the equations ~\eqref{C1} -~\eqref{kovderR} it will give rise to the known functions $\left\langle {R^{p+1}} \right\rangle$. 
\newline \indent If we apply averaging to $R'^{p+1}$, the number of independent functions will be usually decreasing as a consequence of an enlarged isotropy group of the new spacetime $\left\langle  {\cal M}\right\rangle$. We can also obtain an algebra of the Killing vectors~\cite{Karl3}.
\newline \indent In practice there are two goals of averaging - the first is an averaging of the spacetime geometry and the second is an averaging of the Einstein equations. We can see that the left hand side of the Einstein equations (rewritten in the tetrad form when the frame is fixed by the Cartan-Karlhede algorithm) contain the sum of the Cartan scalars and these can be integrated simply as scalar functions. Einstein equations are nonlinear in metric tensor, so we can expect that after averaging we will obtain equations in the form
\begin{equation}
{R^\mu_{\,\,\,\nu}}\left(\overline{g_{\alpha \beta}}\right)-\frac{1}{2}{R}\left(\overline{g_{\alpha \beta}}\right) \delta^\mu_{\,\,\,\nu} +C^\mu_{\,\,\,\nu}=8 \pi T^\mu_{\,\,\,\nu}\left(\overline{g_{\alpha \beta}}\right) .
\end{equation}
Here we suppose that ${R^\mu_{\,\,\,\nu}}\left(\overline{g_{\alpha \beta}}\right)$ is the macroscopic Ricci tensor, which is obtained from the averaged metric $\overline{g_{\alpha \beta}}$. The same holds for $T^\mu_{\,\,\,\nu}\left(\overline{g_{\alpha \beta}}\right)$. In several cases we explicitly suppose the form of the metric structure on the averaged manifold $\overline {\cal M}$ - for example in cosmology it is usual to suppose homogeneous and isotropic FRW models. It is questionable whether this kind of ansatz is adequate. It is straightforward to create perturbations from the symmetric spaces but the inverse procedure is not so clear. By averaging inhomogeneous metric we could also obtain a situation, where the averaged spacetime has nonzero Weyl tensor or where the correlation term is not in the form of a homogeneous and isotropic perfect fluid. It is also ambiguous how to interpret the correlation term. 

First we could use the averaging of Cartan scalars described above and obtain a new macroscopic metric tensor $\left\langle {g_{\alpha \beta}} \right\rangle$ (in general not very simple). Einstein tensor is created from $\left\langle {g_{\alpha \beta}} \right\rangle$. The correct averaging procedure is guaranteed, but the macroscopic metric is gained by a rather difficult method (how to obtain the one-forms $\bomega^i$ from the Cartan scalars $R^{p+1}$ is shown e.g. in~\cite{find}). The correlation term is equal to zero - or more precisely, the geometrical correction is hidden into the macroscopic Ricci tensor ${\left\langle R^\mu_{\,\,\,\nu} \right\rangle}$ ($\left\langle R^\mu_{\,\,\,\nu} \right\rangle$ is constructed using Cartan scalars averaged according to the definition~\eqref{stred}). The advantage of this approach is the possibility to see how the symmetry is increasing after averaging.

More straightforward and for its simplicity more acceptable is the second approach: Suppose the averaged (macroscopic) metric tensor $\overline{g_{\alpha \beta}}$ is given (e.g. spherical symmetric, homogeneous, FRW,...). Then compute the averaged Cartan scalars and compare it with the Cartan scalars for the macroscopic metric - it is possible to see if the form is the same and under which conditions these two are comparable. Now we have two Ricci tensors - the first one is the macroscopic $R^\mu_{\,\,\,\nu}\left(\overline{g_{\alpha \beta}}\right)$ (built from the known $\overline{g_{\alpha \beta}}$) and the second one is $\left\langle R^\mu_{\,\,\,\nu} \right\rangle$ (in the previous paragraph these two were the same). We can define the correlation term as 
\begin{equation}
C^\mu_{\,\,\,\nu}=\left\langle R^\mu_{\,\,\, \nu}\right\rangle - \frac{1}{2}\left\langle R\right\rangle \delta^\mu_{\,\,\, \nu} - {R^\mu_{\,\,\,\nu}}\left(\overline{g_{\alpha \beta}}\right)-\frac{1}{2}{R}\left(\overline{g_{\alpha \beta}}\right) \delta^\mu_{\,\,\,\nu}
\end{equation}
The Ricci tensor ${R^\mu_{\,\,\,\nu}}\left(\overline{g_{\alpha \beta}}\right)$ satisfies contracted Bianchi identities and as a consequence the locally conserved object is not the tensor $T^\mu_{\,\,\,\nu}\left(\overline{g_{\alpha \beta}}\right)$ but the expresion $T^\mu_{\,\,\,\nu}\left(\overline{g_{\alpha \beta}}\right)-C^\mu_{\,\,\,\nu}$. Correlation term can be interpreted as a part of the conserved stress-energy tensor
\begin{equation}
^{(ef)}T^\mu_{\,\,\,\nu}=T^\mu_{\,\,\,\nu}\left(\overline{g_{\alpha \beta}}\right)-C^\mu_{\,\,\,\nu}.
\end{equation}
\indent We can divide averaging into several steps: guess the right macroscopic metric, compute an averaged Cartan scalars and find the correlation term, which can modify the macroscopic metric. 
\newline \indent The question is to decide between these two approaches~\cite{scalars}. In the first one, the procedure is unambiguous and the averaged metric tensor can be constructed (despite technical difficulty). The second one is much easier - it remains to be clarified whether it is possible to use the simplified metric without loosing an important information about the inhomogeneous metric. In cosmology the question is under which circumstances it is possible to characterize the spacetime by only one scale function $a(t)$ and how the form of $a(t)$ is changed by the correlation term. It would cause a problem, if the correlation term did not satisfy the form of stress energy tensor of the ``guessed'' metric (a homogeneous and isotropic perfect fluid in the case of FRW spacetime) and its magnitude would not be negligible. Then we have to use the first approach. 
\newline \indent  Similar situation is present in the theory of Macroscopic Gravity~\cite{Zal1},~\cite{Zal2} - it is necessary to choose which averaged object will be considered as fundamental. In MG the main geometrical objects used in the averaging procedure are Christoffel symbols. In our case, the first possibility is to choose the Riemann tensor (and it's covariant derivatives) because we average Cartan scalars, the second one is the macroscopic metric.
\newline \indent So far, we were dealing with scalars averaged at a single point. If we wanted to obtain a unique prescription for the averaged scalar field, we should have a rule how to choose a domain at the point $x'$ from a given domain at $x$. This problem was discussed by Zalaletdinov in the context of MG~\cite{Zal3}, where the definition of the averaged geometrical objects depends on the choice of the bilocal operators. We will leave this rule unspecified but we will be guided by the symmetries of spacetime. In the next chapter we will assume thick spherical shells for averaging Cartan scalars in an LTB spacetime.
\newline \indent The problem remains how to practically use the constraints~\eqref{vazba}. For doing some explicit calculations, we usually use the fixed frame formalism \cite{find}, where $I^\alpha_{\,\,\,|K}$ correspond to gradients of coordinates and Ricci rotation coefficients $\left\{x^\mu_{\,\,\,|k}, \gamma ^m _{\,\,\,kn}\right\}$ and we have to deal with the difficulty how to average tetrad. In the next chapters we will use the minimal set of Cartan scalars introduced by MacCallum and \AA man~\cite{minimal} and implemented in the algebraic program SHEEP~\cite{Aman}. 
\newline \indent Next, remark should be added. The whole averaging procedure strongly depends on the choice of the frame. In some spacetimes the tetrad can be chosen in a well defined way. This usually works well for spacetimes with an additional symmetry (as will be the case for the spherically symmetric LTB metric discussed in the next section), but the method is not suited e.g. for the general perturbations of FRW, where the frame is restricted only by the algebraic property of spacetime. Another possibility would be to choose the frame by minimizing a certain kind of functional as done by Behrend~\cite{Juliane} in the context of averaging.
\newline \indent Correct averaging should not change the metric structure of the space with a constant curvature. In this case there is only one nonzero Cartan scalar (Ricci scalar or Lambda term in NP formalism), which is constant and the averaging does not change its value. If we have a constant curvature space and perform averaging by Cartan scalars, we obtain the same space.

\section {Cartan scalars of FRW spacetime}
\indent \indent For its simplicity it is most common to use the FRW model as a template for interpreting the cosmological data. It is believed that it is a good approximation of the universe over the large scales. We will consider a flat FRW metric
\begin{equation}
ds^2=-dt^2+a(t)^2(dx^2+dy^2+dz^2).
\end{equation}
Following computations are performed using the algebraic program SHEEP. Nonzero Cartan scalars are
\begin{equation}
\label{FRW1}
\phi_{00'}=\phi_{22'}=2\phi_{11'}=-\frac{1}{2}a^{-1}a_{,tt}+\frac{1}{2}a^{-2}(a_{,t})^2,
\end{equation}
\begin{equation}
\label{FRW2}
\Lambda=\frac{1}{4}a^{-1}a_{,tt}+\frac{1}{4}a^{-2}(a_{,t})^2,
\end{equation}
\begin{eqnarray}
D \phi_{00'}&=&D \phi_{33'}=3D \phi_{11'}=3D \phi_{22'} = -\frac{1}{2\sqrt{2}}a^{-1}a_{,ttt}\nonumber\\
&+& \frac{5}{2\sqrt{2}}a^{-2}a_{,t}a_{,tt}- \sqrt{2}a^{-3}(a_{,t})^3, \label{FRW3}
\end{eqnarray}
\begin{equation}
\label{FRW4}
D\Lambda_{00'}=D\Lambda_{11'} = \frac{1}{4 \sqrt{2}}a^{-1}a_{,ttt} + \frac{1}{4 \sqrt{2}}a^{-2}a_{,t}a_{,tt} 
-\frac{1}{2 \sqrt{2}}a^{-3}(a_{,t})^3.
\end{equation}
\indent Now, if we have an inhomogeneous model, we can compare averaged Cartan scalars with the FRW case. By comparing two different sets of scalars, we can see under which conditions we can obtain an effective FRW metric by averaging.

\section {LTB metric}
\indent \indent The Lema{\^i}tre-Tolman-Bondi (LTB) metric~\cite{Lemaitre},~\cite{Tolman},~\cite{Bondi} is a spherically symmetric exact solution of the Einstein equations. It corresponds to an inhomogeneous dust with the stress energy tensor 
\begin{equation}
T_{\mu \nu}=\rho u_\mu u_\nu, 
\end{equation}
where $u_\mu$ is 4-velocity of a dust with a density $\rho$. For the recent review of LTB metric see e.g.~\cite{Bolejko},~\cite{Hellaby}. The line element reads
\begin{equation}
ds^2=-dt^2+\frac{(R')^2}{1+2E(r)}dr^2+R^2(t,r)(d \theta^2 +\mbox{sin}^2( \theta) d \phi^2),
\end{equation}
where $E(r)$ is an arbitrary function and the prime denotes partial derivative with respect to $r$. Function $R(t,r)$ obeys Einstein equations if
\begin{equation}
\label{LTB}
R_{,t}^2=2E+ \frac{2M}{R} + \frac{\Lambda}{3} R^2.
\end{equation}
$M=M(r)$ is another arbitrary function of integration. The energy density $\rho$ is determined by the equation
\begin{equation}
4 \pi \rho = \frac{M'}{R'R^2}.
\end{equation}
\indent The function $E(r)$ determines a curvature of the space $t=const.$ (which is flat for $E(r)=0$) and the function $M(r)$ is the gravitational mass contained within the comoving spherical shell at any given $r$. Equation~\eqref{LTB} can be integrated to give the result
\begin{equation}
\int\limits_0^R {{{d\tilde R} \over {\sqrt {2E + {{2M} \over {\tilde R}} + {1 \over 3}\Lambda \tilde R^2 } }}}  = t - t_B (r).
\end{equation}
$t_B(r)$ is the third free function of $r$ (called the bang time function). In the LTB model, in general, the Big Bang is not simultaneous as in the FRW case, but it depends on the radial coordinate $r$. The given formulas are invariant under transformation $\tilde{r}=g(r)$. We can use this freedom to choose one of the functions $E(r), M(r)$ and $t_B(r)$. For $\Lambda=0$ the above equation can be solved explicitly - when $E<0$ (elliptic evolution)
\begin{eqnarray}
R\left( {t,r} \right) &=& {M \over {\left( { - 2E} \right)}}\left( {1 - \cos \eta } \right),\nonumber\\
\eta  - \sin \eta  &=& {{\left( { - 2E} \right)^{3/2} } \over M}\left( {t - t_B} \right).
\end{eqnarray}
If E = 0 (parabolic evolution)
\begin{equation}
\label{flatltb}
R\left( {t,r} \right) = \left[ {{9 \over 2}M\left( {t - t_B } \right)^2 } \right]^{1/3}, 
\end{equation}
when $E>0$ (hyperbolic evolution)
\begin{eqnarray}
R\left( {t,r} \right) &=& {M \over {2E}}\left( {\cosh \eta  - 1} \right),\nonumber\\
\sinh \eta  - \eta  &=& {{\left( {2E} \right)^{3/2} } \over M}\left( {t - t_B \left( r \right)} \right).
\end{eqnarray}

\section {Averaging LTB spacetime}
\indent \indent For simplicity we will consider the situation when $E=0$. Unfortunately Cartan scalars for the exact solution listed above are too complicated. We will deal only with an areal function $R(t,r)$. The first guess would be to investigate the separated form $R(t,r)=A(t)B(r)$. However, by the simple radial transformation $dr'=B'(r)dr$ we obtain flat FRW spacetime (the result is easily checked by computing Cartan scalars, which depend only on $t$ coordinate).
\newline \indent Next, we will assume the ansatz
\begin{equation}
\label{WKB}
R\left( {t,r} \right) = A(t,r) \exp{\psi (t,r)},
\end{equation}
where $\psi (t,r)$ is quickly varying function, $\psi << \psi_{,x} \sim \psi_{,xy} \sim \psi_{,xyz}$, where x, y and z denote time or radial coordinate. $\psi_{,x}$ is also much bigger than $A(t,r)$ and its derivatives. In order to compute Cartan scalars we will use the null tetrad 
\begin{eqnarray}
\bomega^0&=&\frac{1}{\sqrt{2}} (dt+R_{,r}dr), \nonumber\\
\bomega^1&=&\frac{1}{\sqrt{2}} (dt-R_{,r}dr), \nonumber\\
\bomega^2&=&\frac{1}{\sqrt{2}} (Rd\theta +iR \sin{\theta} d \phi), \nonumber\\
\bomega^3&=&\frac{1}{\sqrt{2}} (Rd\theta -iR \sin{\theta} d \phi).
\end{eqnarray}
Nontrivial zero-order Cartan scalars are
\begin{equation}
\psi_2= - \frac{1}{6} (R_{,r})^{-1}R_{,ttr} + \frac{1}{6} R^{-1}R_{,t} (R_{,r})^{-1} R_{,tr} + \frac{1}{6} R^{-1}R_{,tt} - \frac{1}{6} R^{-2} (R_{,t})^2,
\end{equation}
\begin{equation}
\phi_{00'}=\phi_{22'}= \frac{1}{2} R^{-1} R_{,t} (R_{,r})^{-1} R_{,tr} - \frac{1}{2} R^{-1} R_{,tt},
\end{equation}
\begin{equation}
\phi_{11'} = - \frac{1}{4} (R_{,r})^{-1} R_{,ttr} + \frac{1}{4} R^{-2} (R_{,t})^2,
\end{equation}
\begin{equation}
\Lambda = \frac{1}{12} (R_{,r})^{-1} R_{,ttr} + \frac{1}{6} R^{-1} R_{,t} (R_{,r})^{-1} R_{,tr} + \frac{1}{6} R^{-1} R_{,tt} + \frac{1}{12} R^{-2} (R_{,t})^{2}. 
\end{equation}
We plug the form~\eqref{WKB} into the spinors. The most important terms are the ones with higher powers of various derivatives of the function $\psi$. Function $A(t,r)$ appears in the same power in numerator and in denominator and is canceled. If we assume the condition $\psi << \psi_{,x} \sim \psi_{,xy} \sim \psi_{,xyz}$, in the leading order all quantities are equal to zero except
\begin{equation}
\Lambda = \frac{1}{2} \psi_{,t}^2.
\end{equation}
\indent Averaging $\Lambda$ over the domain ${\cal D}$ of the shape of the thick shell (times a certain time interval) gives a nonzero contribution which can be constant by a suitable choice of $\psi$ and ${\cal D}$. The first order Cartan scalars contain more terms (higher order Cartan scalars are equal to zero). Lengthy but straightforward calculation shows, that in this approximation they are (in the leading order) all equal to zero. For example the simplest one is
\begin{eqnarray}
D\phi_{00'}&=& \frac{1}{2\sqrt{2}}R^{-1} R_{,t} (R_{,r})^{-1} R_{,ttr}- \frac{3}{2\sqrt{2}} R^{-1} R_{,t} (R_{,r})^{-2} (R_{,tr})^{2} \nonumber\\
&+& \frac{1}{2\sqrt{2}}R^{-1} R_{,t} (R_{,r})^{-2} R_{,ttr} - \frac{1}{2\sqrt{2}}R^{-1} R_{,t} (R_{,r})^{-3} R_{,tr} R_{,rr} \nonumber\\
&-& \frac{1}{2\sqrt{2}}R^{-1} R_{,ttt} + \frac{3}{2\sqrt{2}}R^{-1} (R_{,r})^{-1} R_{,tt} R_{,tr} - \frac{1}{2\sqrt{2}} R^{-1} (R_{,r})^{-1} R_{,ttr}  \nonumber\\
&+& \frac{1}{2\sqrt{2}}R^{-1} (R_{,r})^{-2} (R_{,tr})^{2} - \frac{1}{2\sqrt{2}}R^{-2} (R_{,t})^{2} (R_{,r})^{-1} R_{,tr} \nonumber\\
&+& \frac{1}{2\sqrt{2}}R^{-2} R_{,t} R_{,tt} - \frac{1}{2\sqrt{2}}R^{-2}  R_{,t} (R_{,r})^{-1} R_{,tr} + \frac{1}{2\sqrt{2}}R^{-2} R_{,tt}.
\end{eqnarray}
\indent Now we suppose that the macroscopic metric is a flat FRW spacetime. To have a correct averaging procedure we should have spinors~\eqref{FRW1},~\eqref{FRW3} and~\eqref{FRW4} equal to zero. If we also assume the class of LTB spacetimes (in our approximation given by $\psi (t,r)$) and the domain ${\cal D}$, where the average of $\Lambda$ is constant, these conditions are fulfilled by (anti-)de Sitter space. Correlation term is in the form of a positive cosmological constant, so the averaged LTB spacetime behaves (in the leading order) as an FRW model with a positive cosmological constant - de Sitter spacetime.
\newline \indent In the flat solution without cosmological constant we know the explicit form of $R(t,r)$~\eqref{flatltb}. If we choose the coordinates where the mass function reads $M(r)=\frac{4}{3} \pi M_0 ^4 r^3$ we can relate bang-time function $t_B(r)$ to our ansatz
\begin{equation}
t_b(r)=t- \frac{[A(t,r) \exp{\psi (t,r)}]^{3/2}}{(9/2)^{3/2} \sqrt{4/3 \pi}M_0^2 r^{3/2}},
\end{equation}
that gives (together with our conditions) big restrictions on the form of $A(t,r)$ and $\psi(t,r)$. This requirement could be relaxed if we allow LTB solution with cosmological constant, where the solutions for $R(t,r)$ involve elliptic functions. This does not give us so strict formula for areal function $R(t,r)$ as in the flat LTB spacetime without cosmological constant. Regularity conditions in the origin $r=r_c$, where time derivatives of $R(t,r_c)$ have to be equal to zero, and no shell crossing condition $R'(t,r)\neq 0$  has to be also fulfilled. Another constraints which would be difficult to satisfy are Bianchi identities. 

We can compare our result with a different approach to averaging in LTB spacetime. Paranjape and Singh showed \cite{Paranjape} that in Buchert equations the backreaction term is equal to zero for a general flat LTB metric (which they call marginally bound LTB) - see \cite{Buchert3} for a generalization of this result. We obtained a different result. The first reason is that they used only spatial averaging while we have used a spacetime one. But most importantly different objects were averaged. Paranjape and Singh averaged a subset of Ricci rotation coefficients for orthogonal frame, namely optical scalars. On the other hand, we average all scalars made from Riemann tensor and its covariant derivatives. As already mentioned in \cite{Paranjape} one can expect additional influence coming from objects not considered in averaging procedure. Moreover, the problem of directly comparing these results is rather difficult due to the nonlinear relation between curvature scalars and Ricci rotation coefficients (as can be seen from Newmann-Penrose equations) which would again introduce correlation terms during averaging. Both approaches have their value, the one used in \cite{Paranjape} is better suited for direct cosmological application, but the method presented here takes more effects into account.

\section {Onion LTB model}
\indent \indent As the next example we investigate the onion model used in~\cite{Biswas} by Biswas, Mansouri and Notari who computed the corrections to luminosity distance--redshift relation. It represents spacetime with radial shells of overdense and underdense regions. The curvature of three dimensional spaces is nonzero ($E(r)>0$), so the evolution of LTB model is hyperbolic. For convenience we will use the rescaled function $a(t,r):=\frac{R(t,r)}{r}$ which is suitable for comparison with FRW model. It reads
\begin{equation}
\label{onion}
a(t,r):=\left(\frac{6}{\pi}\right)^{1/3}t^{2/3}(1+ Lt^{2/3} \frac{1}{r} \sin{\pi r} \sin{\pi r})
\end{equation}
If we take the trace of the Einstein equations we will find that Ricci scalar (which is proportional to $\Lambda$ term in NP formalism) behaves in the same way as the matter density (assuming zero cosmological constant and the equation of state $p=(\gamma-1)\rho$). The metric function $a(t,r)$ looks like perturbation of the flat dust FRW spacetime, where density scales like $\rho \propto \frac{1}{a(t)^{3\gamma}}$ and we assume that $L$ is a small parameter. From the form of the metric we demand that the averaged spacetime is Einstein-de Sitter (EdS). Ricci spinor of LTB spacetime is in the form of the perfect fluid. If we perform averaging the condition for the Ricci spinor to describe perfect fluid does not change. Weyl spinor and higher order Cartan scalars will be discussed later. The most important Cartan scalar is $\Lambda$ term which reads
\begin{eqnarray}
\Lambda(r,t,L)&=&\frac{1}{12} \frac{1}{a^2(a+a_{,r})r} [ 3a^2a_{,tt} + a^2r a_{,ttr} + 3a(a_{,t})^2 \nonumber\\
&+& 2a_{,tt}aa_{,r}r+2a_{,t}raa_{,tr} +(a_{,t})^2ra_{,r} + a_{,r}rK + 3Ka + aK_{,r} ].
\end{eqnarray}
Function $K(r,L)$ is related to curvature function $E(r)$ by
\begin{equation}
K(r,L)=- \frac{2E(r)}{r^2}= \frac{-L}{\pi r} \sin{\pi r} \sin{\pi r}
\end{equation}

Here we perform averaging on the constant time surface. We choose one point and the domain $\Omega$ and denote a new averaged function $\left\langle \Lambda \right\rangle$ which is only time dependent. Next, we expand an averaged $\Lambda$ term in powers of $L$ and we obtain series that looks like
\begin{equation}
\label{lambda}
\left\langle \Lambda \right\rangle \approx \frac{A}{t^{2}} + \frac{B}{t^{4/3}}L + \frac{C}{t^{2/3}}L^2 + \frac{D}{t^{0}}L^3
\end{equation}
The coefficients in front of different powers of $L$ depend on the chosen point and the domain $\Omega$ and can be calculated as follows. In the definition of the average value of $\Lambda$~\eqref{stred} (but here ${\cal D}$ denotes three dimensional surface), we expand in powers of $L$ the integrated expressions in numerator and denominator separately and compute coefficients in front of the time-dependent terms. Then we expand the whole expression and obtain equation~\eqref{lambda}. Now, we have an averaged EdS background, so the scale factor $a(t)$ is proportional to $t^{2/3}$ and the density scales like $\rho \propto \frac{1}{t^{2\gamma}}$. Now let us assume that we can use this expression for the additional terms in $\left\langle \Lambda \right\rangle$ that deviate from EdS. The dominant term describes the dust as expected. Expression proportional to $L$ has an equation of state $p=-\frac{1}{3} \rho$ and behaves like curvature (as interpreted in \cite{sol1}). Next term can already cause acceleration with the dependence of density on pressure $p=-\frac{2}{3} \rho$ and the term proportional to $L^3$ behaves like cosmological constant. 
\newline \indent We can play the same game with nonzero Weyl scalar $\psi_2$ and we can see that the first nonzero contribution to $\left\langle \psi_2 \right\rangle$ is proportional to $L$. In order to obtain EdS background, we need to have $\left\langle \psi_2 \right\rangle=0$. Also all higher order Cartan scalars should be comparable with the averaged spacetime up to the corrections in powers of L.

\section {Linearized gravitational wave}
\indent \indent In the last simple example, we will show the non-triviality of averaging. We assume monochromatic linearized gravitational wave with selected polarization propagating in the direction $z$ on the Minkowski background 
\begin{equation}
ds^2=-dt^2+\left(1+A\mbox{sin}(t-z) \right)dx^2+(1-A\mbox{sin}(t-z))dy^2+dz^2.
\end{equation}
$A$ is a small parameter describing the amplitude of the gravitational wave. Non zero lowest-order Cartan scalars read 
\begin{eqnarray}
\Psi_4&=&\left(\frac{1}{2} \mbox{sin}(t) \mbox{cos}(z)-\frac{1}{2}\mbox{sin}(z) \mbox{cos}(t)\right)A+O(A^3), \nonumber\\
\Phi_{22}&=&\left(\frac{1}{4}-\frac{3}{4}\mbox{cos}(z)^2-\frac{3}{4}\mbox{cos}(t)^2+\frac{3}{2}\mbox{cos}(t)^2\mbox{cos}(z)^2 \right)A^2 \nonumber\\
&+&\left(\frac{3}{2}\mbox{sin}(t) \mbox{cos}(z) \mbox{sin}(z)\mbox{cos}(t)\right)A^2+O(A^4).
\end{eqnarray}
If we integrate over several wave lengths, Weyl scalar vanishes and nonzero Ricci scalar $\Phi_{22}$ is constant. Now we assume that averaged spacetime is Minkowski background. $\Phi_{22}$ can be put into the right hand side of the Einstein equation and interpreted as the correlation term which behaves like a null fluid and serves as an effective stress energy tensor of the gravitational wave. If we would like to determine its influence on the background we would need to consider next-order Cartan scalars.

\section {Conclusion}
\indent \indent Theory of Cartan scalars is commonly used for equivalence problem. We have applied this theory in the context of averaging in GR and cosmology. There are two different ways how to perform averaging. In the first one the correlation term is equal to zero, but the averaged geometry is explicitly constructed. In the second approach we assume the form of the smooth metric tensor and compute correlation term. We used the second approach for computation of backreaction in two different LTB models. Correlation term behaves as a cosmological constant in the first example and the curvature term plus small terms causing acceleration in the second example. Thus inhomogeneity of spacetime may serve as a reason for accelerated expansion when viewed in averaged picture of standard cosmological models. This is in contrast with the solutions of \cite{sol1} and \cite{scalars}  where correlation term behaves as a curvature term and does not lead to acceleration. We have also shown the non-triviality of averaging in the case of monochromatic linearized gravitational wave.

\section*{Acknowledgments}
We would like to thank M. Bradley, A. A. Coley and J. Hru\v ska for useful discussions. We would also like to thank J.~E. \AA man and M. Bradley for providing us with the algebraic program SHEEP.  P.K. was supported by grants GAUK 398911 and SVV-267301. O.S. was supported by grant GA\v{C}R 14-37086G.

\end {document}